\documentclass[11pt,twoside]{article}
\usepackage{asp2010}
\usepackage{graphicx}
\usepackage{marvosym}
\usepackage{natbib}
\bibpunct{(}{)}{;}{a}{}{,}

\resetcounters

\newcounter{Rco}

\newcommand{\logg}{\mbox{$\log g$}}
\newcommand{\loggw}[1]{\mbox{$\log g\hspace{-0.5mm} =\hspace{-0.5mm}  #1$}}
\newcommand{\mc}[3]{\multicolumn{#1}{#2}{#3}}
\newcommand{\kK}{\mbox{\rm kK}}

\newcommand{\sla}{\raisebox{-0.10em}{$\stackrel{<}{{\mbox{\tiny $\sim$}}}$}}
\newcommand{\spm}{\mbox{\raisebox{0.20em}{{\tiny \hspace{0.2mm}\mbox{$\pm$}\hspace{0.2mm}}}}}

\newcommand{\Teff}{\mbox{$T_\mathrm{eff}$}}

\newcommand{\Teffw}[1]{\mbox{$\Teff\hspace{-0.5mm}=\hspace{-0.5mm}#1\,\mathrm{K}$}}

\newcommand{\aador}{AA\,Dor}

\def\tmap{\emph{TMAP}}
\def\tmad{\emph{TMAD}}

\begin{document}

\title{Recent Investigations on AA \,Doradus}
\author{Thomas Rauch
\affil{Institute for Astronomy and Astrophysics,
       Kepler Center for Astro and Particle Physics,
       Eberhard Karls University,
       Sand 1,
       D-72076 T\"ubingen,
       Germany\\email: rauch@astro.uni-tuebingen.de}
}
\begin{abstract}
AA\,Dor is an eclipsing, post common-envelope binary with an 
sdOB-type primary and an unseen low-mass secondary, believed to
be a brown dwarf. Eleven years ago, a NLTE spectral analysis 
of the primary showed a discrepancy with the surface gravity 
that was derived by radial-velocity and light-curve analysis
that could not be explained.

Since then, emission lines of the secondary were identified in
optical spectra and its orbital-velocity amplitude was measured.
Thus, the masses of both components are known, however, within
relatively large error ranges. The secondary's mass was found
to be around the stellar hydrogen-burning mass limit and, thus,
it may be a brown dwarf or a late M-type dwarf. In addition, a 
precise determination of the primary's rotational velocity 
showed recently, that it rotates at about 65\,\% of bound
rotation -- much slower than previously assumed.

A new spectral analysis by means of metal-line blanketed, 
state-of-the-art, non-LTE model atmospheres solves the so-called 
gravity problem in AA\,Dor -- our result for the surface gravity 
is, within the error limits, in agreement with the value from 
light-curve analysis.

We present details of our recent investigations on AA\,Dor.
\end{abstract}

\section{Introduction}
\label{sect:introduction}

\aador\ is a close, eclipsing binary \citep{kilkennyhilditchpenfold_1978}
and, thus, it is possible to determine its geometrical parameters
(inclination $i$, radii $r_\mathrm{pri}$ and $r_\mathrm{sec}$) precisely
by high-speed photometry. This has been done now over a
time-span of about 50\,000 eclipses and orbital period
$P$ = 0.261\,539\,7363 (4)\,d \citep{kilkenny_2011} 
and 
$i=89\fdg 21\pm 0\fdg 30$ \citep{hilditch_2003}
were determined.

An early model of \aador\ \citep{paczynski_1980}
based on a light-curve analysis assumed an sdOB-type primary with
$M_\mathrm{pri} = 0.36\,\mathrm{M_\odot}$ 
and a low-mass secondary with
$M_\mathrm{sec} = 0.054\,\mathrm{M_\odot}$.
Paczy\'nski estimated that 
the common envelope was ejected $5\cdot10^5$ years ago and 
the primary will become a degenerate, hot white dwarf in another $5\cdot10^5$ years.
Within $5\cdot10^{10}$ years then, gravitational radiation
will reduce the orbital period to about 38 minutes and the 
degenerate secondary will overflow its Roche lobe,
making \aador\ a cataclysmic variable.
However, since \aador\ is a PCEB and is now a pre-cataclysmic variable, 
it is a key object for understanding such an evolution. 
Moreover, the low-mass companion, just at the limit of the hydrogen-burning mass, 
and its angular momentum are important for the understanding of the common-envelope ejection mechanism
\citep{liviosoker_1984}.

\citet{paczynski_1980} already pointed out the importance to find signatures
of the irradiated secondary to measure its radial-velocity curve
in order to refine the model of \aador. 
Quite recently,
\citet{vuckovicetal_2008} found spectral lines of the secondary in the 
UVES\footnote{Ultraviolet and Visual Echelle Spectrograph at ESO's VLT KUEYEN} 
spectra (ProgId 66.D$-$180)
and measured its 
orbital-velocity amplitude ($K_\mathrm{sec} > 230\,\mathrm{km/sec}$), 
both components' masses are known now
$M_\mathrm{pri} = 0.45\,\mathrm{M_\odot}$ and
$M_\mathrm{sec} = 0.076\,\mathrm{M_\odot}$, 
albeit with large error bars. 

Based on their spectral analysis of the sdOB primary,
\citet{klepprauch_2011} found
$M_\mathrm{pri} = 0.4714\pm 0.0050\,\mathrm{M_\odot}$ and
$M_\mathrm{sec} = 0.0725 - 0.0863\,\mathrm{M_\odot}$
within small error limits.
The scenario \citep{trimbleaschwanden_2001}
\begin{quote}
``that a planet belonging to \aador\ tried to swallow its star
during common-envelope binary evolution, rather than the converse
\citep{rauch_2000}''
\end{quote}
($M_\mathrm{sec} = 0.066\,\mathrm{M_\odot}
                    \approx 70\,\mathrm{M\raisebox{-1mm}{\small \Jupiter}}$)
is not valid anymore.

The results of \citet{klepprauch_2011} for the primary,
\Teffw{42000\pm 1000} and \loggw{5.46\pm 0.05}, 
are now in good agreement with the photometric model of \citet{hilditch_2003}. 
Detailed information about \aador\ and previous analyses 
are given in 
\citet{rauch_2000, rauch_2004}, 
\citet{fleigetal_2008}, and 
\citet{klepprauch_2011}.

We will not repeat the description of the \citet{klepprauch_2011}
analysis here but we will emphasize problems that we encountered
during that analysis.

\section{Atmosphere Modeling and Spectral Analysis of sd(O)B-type stars}
\label{sect:atmospheres}

\begin{figure}[ht]
\setlength{\textwidth}{13.85cm}
\plotone{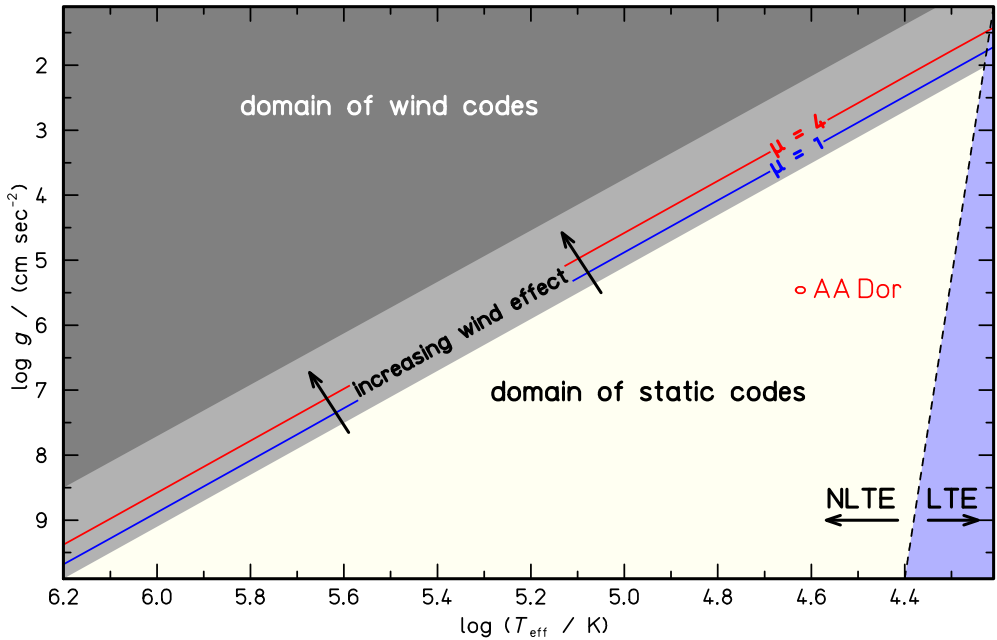}
\caption{Position of the primary of \aador\ in the \logg\ - \Teff\ plane.
         Its symbol's size represents the errors of \citet{klepprauch_2011}.
         The $\mu = 1$ and $\mu = 4$ lines indicate the Eddington limits for
         pure hydrogen and helium atmospheres, respectively.}
\label{fig:domains}
\end{figure}

\citet{muelleretal_2010} presented a spectral analysis of the
primary of \aador\ that resulted in a too-low \Teff. This initiated the
re-analysis of \citet{klepprauch_2011}. They used the
T\"ubingen Model-Atmosphere Package 
(\tmap\footnote{http://astro.uni-tuebingen.de/\raisebox{.3em}{$\sim$}TMAP})
to calculate plane-parallel NLTE models in hydrostatic and radiative equilibrium.
Fig.~\ref{fig:domains} shows that the primary of \aador\ is located in
the domain of static NLTE models. The limit between NLTE and LTE shown in 
Fig.~\ref{fig:domains} is not stringent. There are, however, deviations from
LTE in any star, at least at high energies and high resolution.
(cf. Jeffery, Pereira, Naslim, \& Behara these proceedings.)

In addition to the necessity to use NLTE models for an appropriate analysis
of the primary of \aador, fully metal-line blanketed model atmospheres
should be used to avoid the so-called Balmer-line problem 
\citep[cf\@.][]{napiwotzkirauch_1994,rauch_2000}.
H+He and the intermediate-mass elements
C+N+O+Mg+Si+P+S were considered using classical model atoms, 
while for Ca+Sc+Ti+V+Cr+Mn+Fe+Co+Ni, the so-called iron-group elements,  
a statistical approach \citep{rauchdeetjen_2003} is employed to
create model atoms.
The most recent atomic data (\tmad\footnote{http://astro.uni-tuebingen.de/\raisebox{.3em}{$\sim$}TMAD})
were used.
530 levels are treated in NLTE with 771 individual lines (from H\,-\,S) and 
19\,957\,605 lines of Ca\,-\,Ni from Kurucz' line lists \citep{kurucz_2009} combined to 636 superlines.
For all models, we use the same model atoms and the same frequency grid.

Two model-atmosphere grids were calculated, firstly a coarse grid (638 models)
within \Teffw{35\,000 - 49\,000} ($\Delta$\,\Teffw{500}) and 
\loggw{5.15 - 6.20} ($\Delta$\,\loggw{0.05}). Secondly,
a finer sub-grid (527 models) with
\Teffw{39\,500 - 43\,500} ($\Delta$\,\Teffw{250}) and 
\loggw{5.30 - 5.60} ($\Delta$\,\loggw{0.01}).
All calculations were performed on computational resources of the 
bwGRiD\footnote{bwGRiD (http://www.bw-grid.de), member of the German D-Grid initiative,
      funded by the Ministry for Education and Research (Bundesministerium f\"ur
      Bildung und Forschung) and the Ministry for Science, Research and Arts
      Baden-W\"urttemberg (Ministerium f\"ur Wissenschaft, Forschung und Kunst
      Baden-W\"urttemberg)}.

\begin{figure}[ht]
\setlength{\textwidth}{13.85cm}
\plottwo{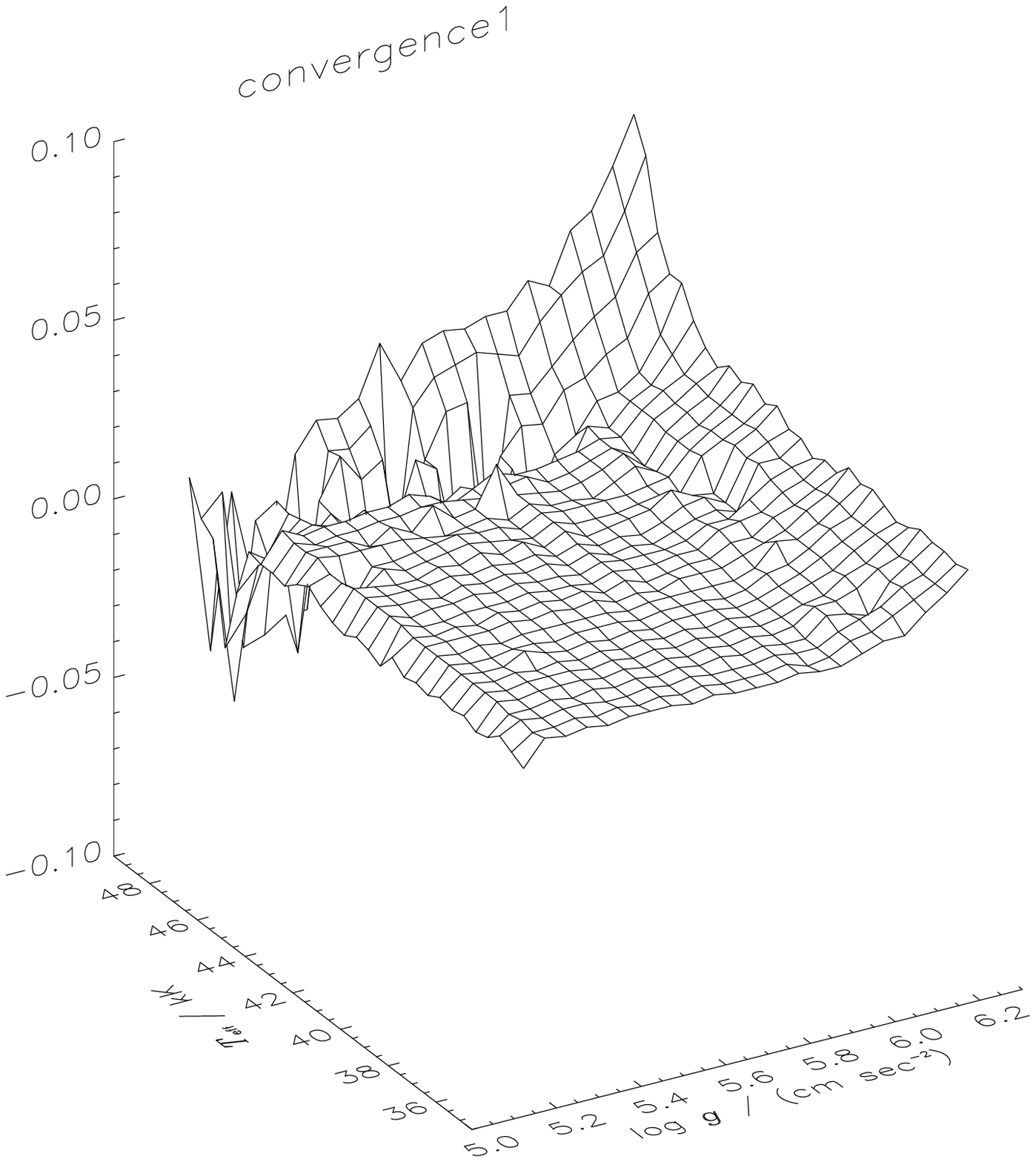}{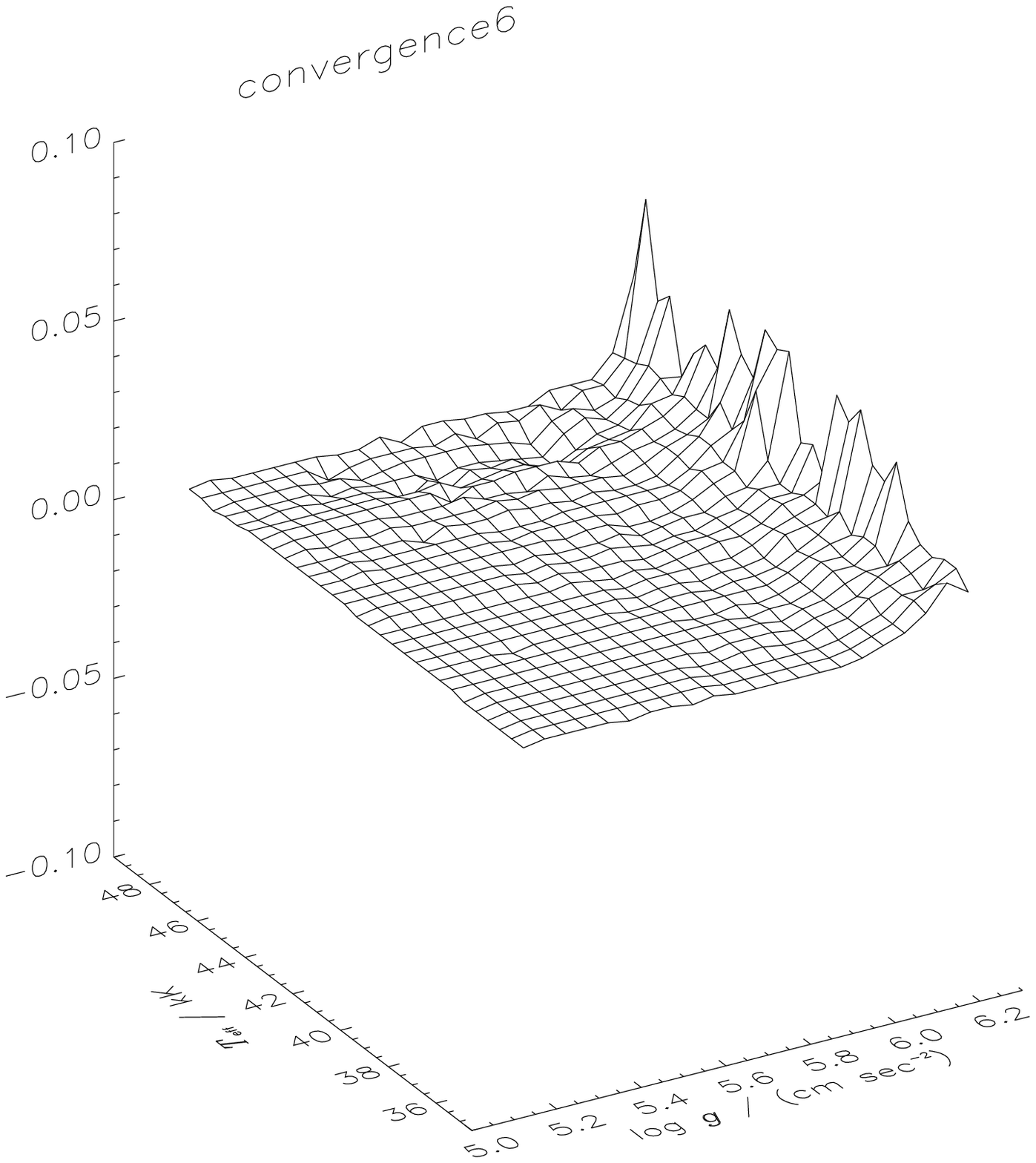}\vspace{5mm}
\caption{$\log (T_\mathrm{eff}^\mathrm{\,model~actual} / T_\mathrm{eff}^\mathrm{\,model~target})$
         at two points in time during the model-grid calculation.}
\label{fig:status}
\end{figure}

The computational time for single models to converge is relatively long, 
and it took more than a diploma student's time to complete the grids.
During their calculation, we performed quality control to check
the grids' status. Fig.~\ref{fig:status} shows a comparison of actual
and desired \Teff\ for two selected phases of model-atmosphere calculation.
The left panel shows the situation just after a grid extension towards higher
\Teff\ and \logg. The right panel shows a later phase where the complete
grid (but the highest \logg) is more homogeneously converged.

\begin{figure}[ht]
\setlength{\textwidth}{13.85cm}
\plotone{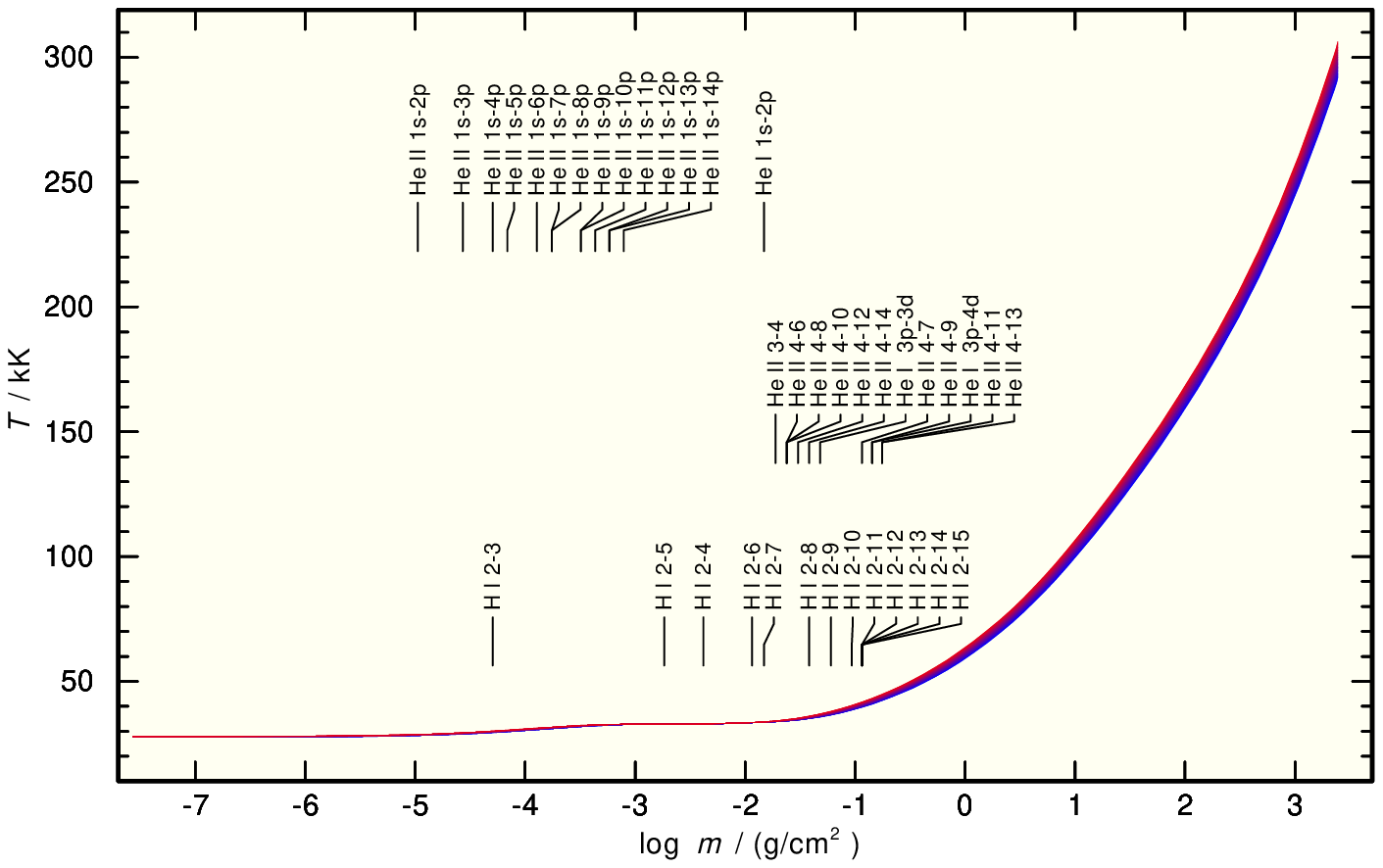}
\caption{Temperature structure of our \Teffw{42\,000} models for all
         \loggw{5.30 - 5.60} ($\Delta$\,\loggw{0.01}). (\loggw{5.30} is blue and \loggw{5.60}
         is red in the online version.) The depth points (formation depths) where
         the line centers of
         selected hydrogen and helium lines 
         become optically thin ($\tau < 1$)
         in the outer atmosphere are marked.}
\label{fig:temperature}
\end{figure}

\begin{figure}[ht]
\setlength{\textwidth}{13.85cm}
\plotone{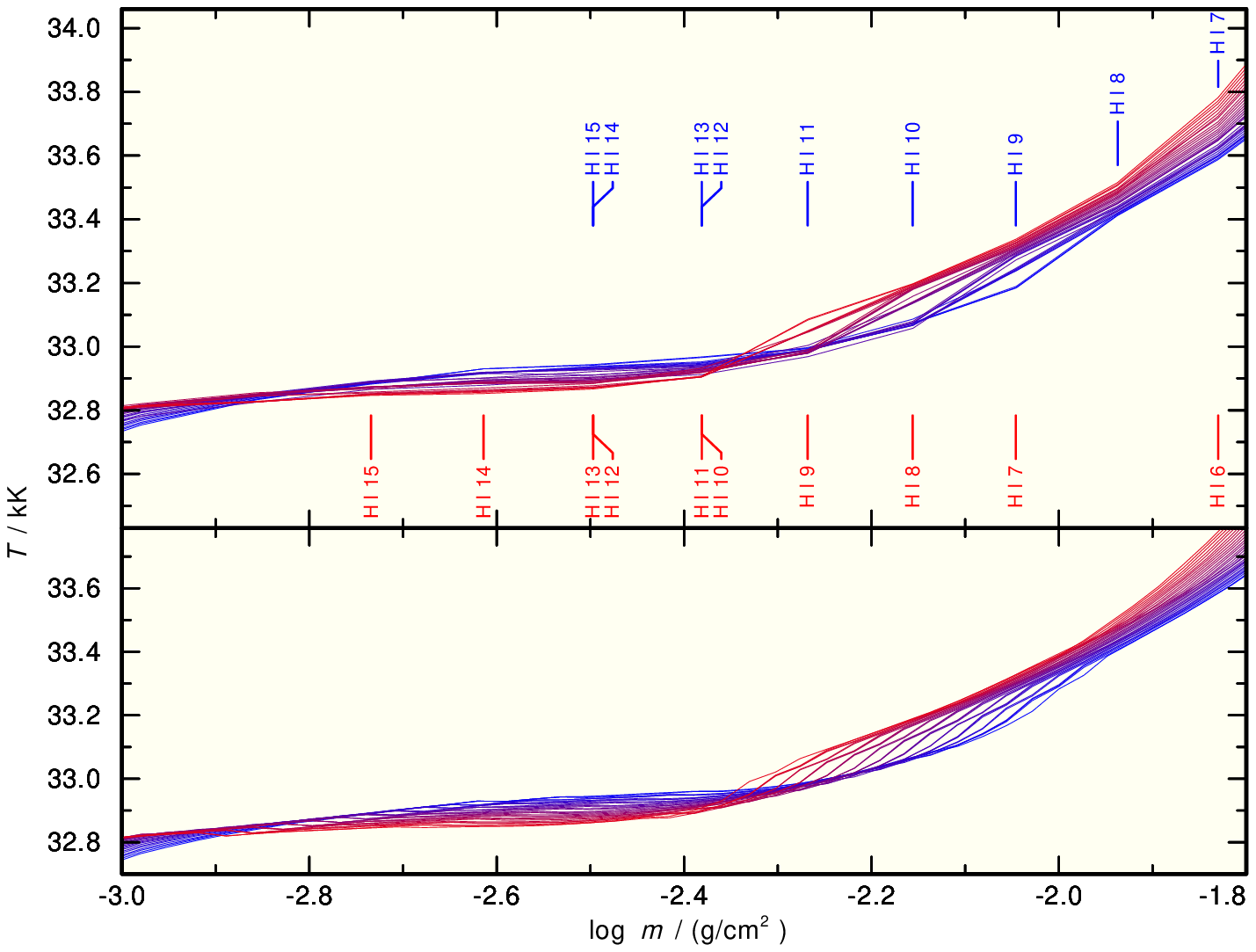}\vspace{-2mm}
\caption{Top panel: Detail of Fig.\,\ref{fig:temperature}. The formation depths of the
         hydrogen absorption edges are marked at top (for \loggw{5.30}, blue) and 
         bottom (\loggw{5.60}, red). Bottom panel: Same as top panel but with a four times
         refined depth-point discretization in the model.
         }
\label{fig:depthpoints}
\end{figure}

It turned out that the temperature structures of the models (Fig.~\ref{fig:temperature}) suggest
that the models are converged much earlier than according to the \Teff\ criterion.
However, even at the converged state, a close look to the temperature
structures of the models reveals a surprise. Fig.~\ref{fig:depthpoints} shows that there is
obviously some grouping in the temperature structures within $-2.9\, \sla\, \log m\, \sla\, -1.9$.
The reason is that the formation depths of absorption edges and lines are dependent on \Teff\ and
\logg. This is demonstrated for \logg\ in Fig.\,\ref{fig:depthpoints}. E.g\@. the absorption
edge of H\,{\sc i} $n=9$ forms at $\log m \approx -2.05$ in the \loggw{5.30} model
and at $\log m \approx -2.27$ in the \loggw{5.60} model. (The individual edge or line that
is responsible for the grouping is not investigated.)
Our \emph{TMAP} model atmospheres consider by default 90 depth points. 
In a test calculation, we used 360 depth points (Fig.\,\ref{fig:depthpoints}) and could
clearly show that the depth-point discretization is the reason of the temperature
grouping.

A higher number of depth points increases the calculation time linearly but
has no significant impact on the analysis. However, one has to be aware that
there may be artificial temperature jumps of a few hundred K in the
line-forming regions of the atmosphere due to numerical 
limitations in the model-atmosphere calculations. This is especially important
in case that $\chi^2$ fits are presented, with the extraordinary small
statistical errors.

\section{The Balmer-Line Problem -- Still present?}
\label{sect:blp}

\begin{figure}[ht]
\setlength{\textwidth}{13.85cm}
\plotone{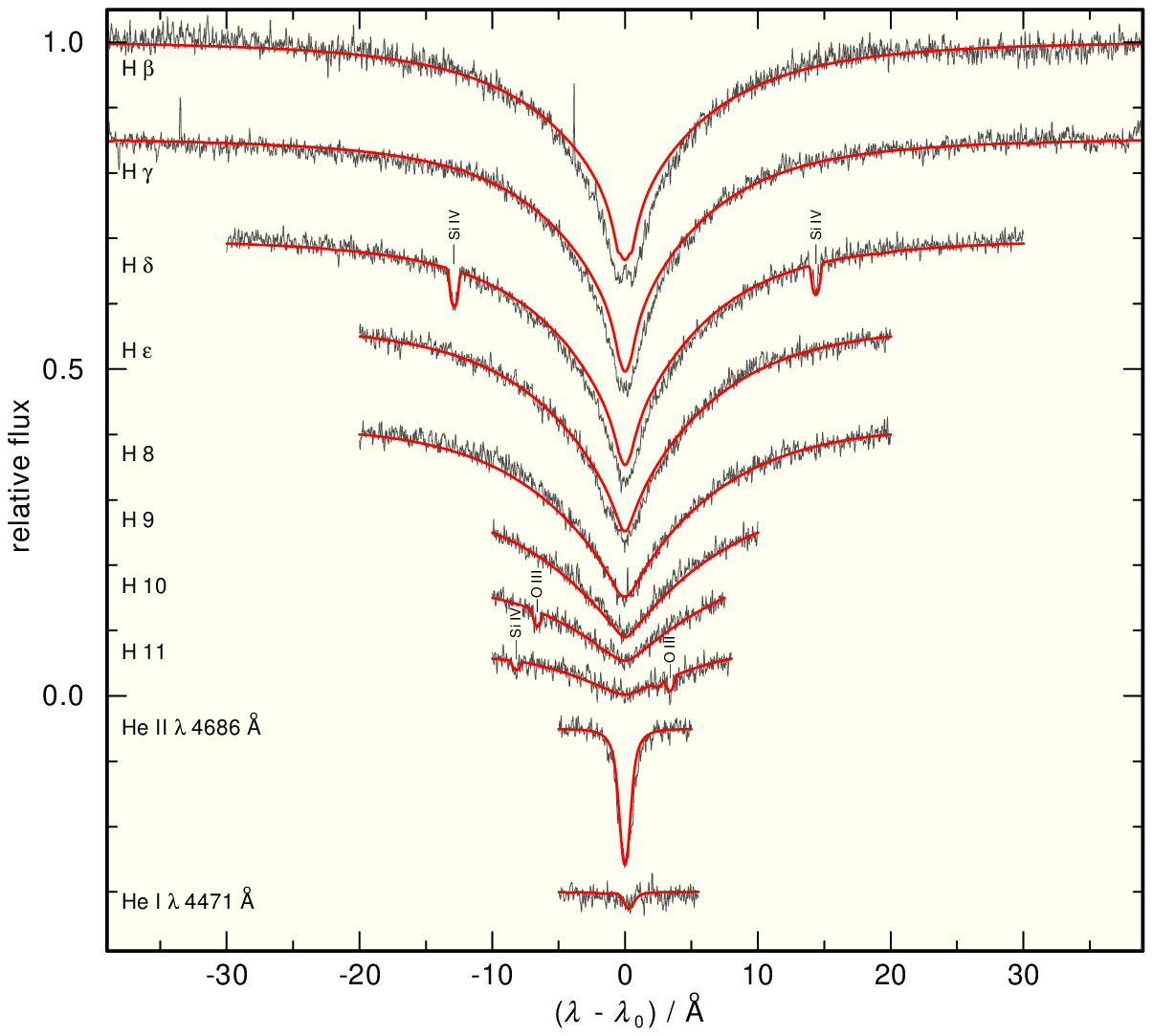}
\caption{Comparison of theoretical line profiles of H and He lines calculated from
         the final \Teffw{42000} and \loggw{5.46} model with the UVES observation
         \citep[for details, see][]{klepprauch_2011}.
         }
\label{fig:blp}
\end{figure}

The final Balmer-line fit (Fig.\,~\ref{fig:blp}) of \citet{klepprauch_2011}, got a comment by
the A\&A editor Ralf Napiwotzki:
\begin{quote}
``It is slightly depressing that after all this improvements and on this high
level of sophistication, the Balmer lines are still not well fitted.''
\end{quote}

\noindent
Indeed, the line cores as well as inner line wings of H\,$\beta$ - $\delta$ are
not well reproduced. Their line centers form at
$-2.4\, \sla\, \log m\, \sla\, -1.8$ (Fig.\,~\ref{fig:temperature}) where
the temperature jumps (Fig.\,~\ref{fig:depthpoints}) occur. A higher
number of depths points, however, did not improve the agreement.

The reason is more likely due to the Balmer-line problem \citep{napiwotzkirauch_1994},
i.e\@. that
opacities are still missing in our
model-atmosphere calculations.
These are 
on the one hand of
chemical species that are as yet unconsidered and
on the other hand of 
species whose abundance is determined too-low.

A new grid of models is already calculating, it will be used to further
investigate this issue.

\section{Conclusions}
\label{sect:conclusions}

\begin{figure}[ht]
\setlength{\textwidth}{10cm}
\plotone{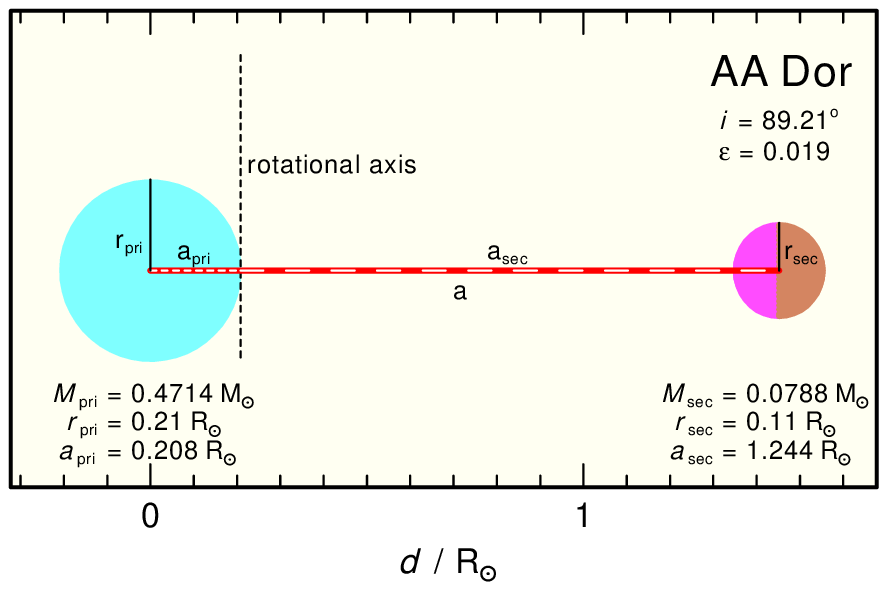}
\caption{Dimension of \aador.}
\label{fig:sketch}
\end{figure}

The sdOB primary of \aador\ is well studied and most of its parameters are determined within
small error ranges. Additional opacities may improve the model atmospheres and e.g\@.
the agreement of synthetic Balmer-line profiles with the observation. 
We summarize parameters of \aador\ in Table\,\ref{tab:para}
and show a sketch of the binary in Fig\@.\,\ref{fig:sketch}.

The masses of both components of \aador\ are well known,
$M_\mathrm{pri} = 0.4714\pm 0.0050\,\mathrm{M_\odot}$ and
$M_\mathrm{sec} = 0.0725 - 0.0863\,\mathrm{M_\odot}$
\citep{klepprauch_2011}.
Since the hydrogen-burning mass limit is $\approx 0.075\,\mathrm{M_\odot}$ 
\citep{chabrierbaraffe_1997,chabrieretal_2000}, it is still unclear
whether the secondary is a brown dwarf or a late M-type dwarf.
In order to make progress and to investigate on the nature of the secondary of
\aador, phase-dependent spectra of \aador\ in the infrared are highly desirable
in order to identify the secondary's spectral contribution and to follow
its spectral evolution during its orbit.

\section{An Answer to an Important Question}
\label{sect:question}
At the very end of the question time following my talk,
Darragh O'Donoghue asked \vspace{-0.5mm}
\begin{quote}
``Thomas, is it right, that with all your spectral analysis,
you arrived at the same result that we got already 30 years ago from
light-curve analysis?''
\end{quote}\vspace{-0.5mm}
Of course, he meant the SAAO group and \logg\ of the primary --
and thus, the $\log g$ problem \citep{rauch_2000},
that is solved now. 
The simple answer to Darragh's question is yes. However, I would like to
point out that -- thanks to the comments of Ron Hilditch, who was referee of
\citet{rauch_2000} -- we took the challenge to solve the \logg\ problem
and finally achieved agreement. This is a good example that, if carried out
with care, theory works. The independent method to determine
\logg\ from light-curve analysis provided a crucial constraint to test
our model atmospheres. Within this course, we improved our
models, the atomic data, as well as our observational, spectral data base 
\citep{rauchwerner_2003}. The errors in \Teff, \logg, and abundances
\citep{fleigetal_2008} for the primary of \aador\ were reduced.

In total, our picture of \aador\ improved, making this one of the best-analyzed PCEB.
We are still not tired to improve the primary's model, but it appears time
to focus now on the outstanding question of the nature of the low-mass companion.
Phase-dependent, high-S/N infrared spectroscopy, that covers the complete orbital period of \aador,
will provide answers.

\clearpage

\begin{table}[p]\centering
\caption{Parameters of \aador\ compiled from 
\citet{rauch_2000}, 
\citet{fleigetal_2008}, 
\citet{muelleretal_2010},
and \citet{klepprauch_2011}.
$\log \varepsilon_\mathrm{X}$ are normalized to $\log \sum\mu_\mathrm{X}\varepsilon_\mathrm{X} = 12.15$.
[X] denotes log(abundance/solar abundance) of species X.
The solar abundances are given by \citet{asplundetal_2009}.}
\label{tab:para}
\vspace{5mm}
\begin{tabular}{r@{\,/\,}lr@{.}llr@{.}lr@{.}l}
\hline
\hline
\noalign{\smallskip}                                                                                
\mc{2}{l}{{\bf Primary}} & \multicolumn{2}{c}{~} & \hspace{50.5mm}\hbox{~}                            \\
\hline                                                                                              
\noalign{\smallskip}                                                                                
\Teff & \kK                                   &  42&0        & \spm 1                               \\
$\log\,(g$& $\mathrm{\frac{cm}{s^2}})$        &  5&46        & \spm 0.05                            \\
\noalign{\smallskip}                                                                                
\hline                                                                                              
\end{tabular}

\begin{tabular}{lllr@{.}lr@{.}l}
           & mass       & number  \\
abundances & fraction   & fraction   & \multicolumn{2}{c}{$\log \varepsilon_\mathrm{X}$} & \multicolumn{2}{c}{[X]}\\           
\hline                                                                                              
\noalign{\smallskip}                                                                                
H          & 9.94E-01 & 9.99E-01 & 12&144          &    0&130 \\
He         & 2.69E-03 & 6.80E-04 &  8&977          & $-$1&967 \\
C          & 1.78E-05 & 1.50E-06 &  6&320          & $-$2&124 \\
N          & 4.15E-05 & 3.00E-06 &  6&621          & $-$1&223 \\
O          & 1.01E-03 & 6.39E-05 &  7&950          & $-$0&754 \\
Mg         & 4.08E-04 & 1.70E-05 &  7&375          & $-$0&240 \\
Si         & 3.05E-04 & 1.10E-05 &  7&186          & $-$0&339 \\
P          & 5.20E-06 & 1.70E-07 &  5&375          & $-$0&050 \\
S          & 3.24E-06 & 1.02E-07 &  5&155          & $-$1&980 \\
Ca         & 5.99E-05 & 1.52E-06 &  6&325          & $-$0&030 \\
Sc         & 3.69E-08 & 8.33E-10 &  3&065          & $-$0&100 \\
Ti         & 2.78E-06 & 5.89E-08 &  4&915          & $-$0&050 \\
V          & 3.73E-07 & 7.42E-09 &  4&015          &    0&070 \\
Cr         & 1.66E-05 & 3.24E-07 &  5&655          &    0&001 \\
Mn         & 9.88E-06 & 1.82E-07 &  5&405          & $-$0&040 \\
Fe         & 1.15E-03 & 2.09E-05 &  7&465          & $-$0&050 \\
Co         & 3.59E-06 & 6.17E-08 &  4&935          & $-$0&069 \\
Ni         & 3.48E-04 & 6.01E-06 &  6&923          &    0&689 \\
\hline                                                                                              
\end{tabular}
\begin{tabular}{r@{\,/\,}lr@{.}llr@{.}lr@{.}l}
\noalign{\smallskip}                                                                                
$M_\mathrm{pri}$& $\mathrm{M_\odot}$          &   0&4714       & \spm 0.0050                        \\
\noalign{\smallskip}                                                                                
$R_\mathrm{pri}$& $\mathrm{R_\odot}$          &   0&21         & $^{+0.028}_{-0.024}$               \\
\noalign{\smallskip}                                                                                
$a_\mathrm{pri}$& $\mathrm{R_\odot}$          &   0&208        & $^{+0.030}_{-0.018}$               \\
\noalign{\smallskip}                                                                                
$L_\mathrm{pri}$& $\mathrm{L_\odot}$          & \mc{2}{c}{120} & $^{+15}_{-20}$                     \\
$v_\mathrm{rot}$& $\mathrm{\frac{km}{sec}}$   & \mc{2}{c}{30}  & \spm 1                             \\
\noalign{\smallskip}                                                                                
\hline                                                                                              
\noalign{\smallskip}                                                                                
\mc{2}{l}{{\bf Secondary}}  & \multicolumn{2}{c}{~} & \hspace{46.5mm}\hbox{~}                       \\
\hline                                                                                              
\noalign{\smallskip}                                                                                
$M_\mathrm{sec}$& $\mathrm{M_\odot}$          &   0&0788       & $^{+0.0075}_{-0.0063}$             \\
\noalign{\smallskip}                                                                                
$R_\mathrm{sec}$& $\mathrm{R_\odot}$          &   0&11         & $^{+0.015}_{-0.013}$               \\
\noalign{\smallskip}                                                                                
$a_\mathrm{sec}$& $\mathrm{R_\odot}$          &   1&244        & $^{+0.181}_{-0.110}$               \\
\noalign{\smallskip}                                                                                
\hline                                                                                              
\noalign{\smallskip}                                                                                
$d$             & $\mathrm{pc}$               & \mc{2}{c}{352} & $^{+20}_{-23}$                     \\
$n_\mathrm{H}$  & $\mathrm{cm^2}$             & \mc{2}{c}{$2\cdot 10^{20}$} & \spm $1\cdot 10^{20}$ \\
\mc{2}{c}{$E_\mathrm{B-V}$}                   &  0&01          & \spm 0.01                          \\
\hline
\end{tabular}
\end{table}

\clearpage

\acknowledgements 
TR is supported by the German Aerospace Center (DLR, grant 05\,OR\,0806).
The attendance of the sdOB5 meeting was funded from a DAAD grant.
The UVES spectra used in this analysis were obtained as part of an ESO Service Mode run,
proposal 66.D-1800.
We gratefully thank the bwGRiD$^4$ project for the computational resources.
This research made use of the SIMBAD Astronomical Database, operated at the CDS, Strasbourg, France.

\vfill

\bibliography{rauch}

\clearpage

\begin{figure}[ht]
\setlength{\textwidth}{10.00cm}
\plotone{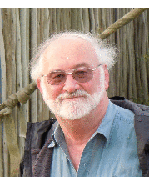}
\end{figure} 
\begin{center}
David Kilkenny, with small tribal face painting, at Spier wine estate, July 27, 2011.
\end{center}
\vspace{1cm}

\noindent
sdOB5 marked the 65$^\mathrm{th}$ birthday of David Kilkenny, who discovered 
that \aador\ is a short-period, eclipsing binary system. Since December 1974,
he observed this system once in a while, covering now a period of more than 46\,119 eclipses, 
and could show that the period increase or decrease is less that 10$^{-14}$\,d/orbit.
\vspace{2mm}

All the best wishes to you, Dave, and have the energy for the next
tens of thousands of eclipses of \aador!

\end{document}